\documentclass[superscriptaddress,amsmath,amssymb,aps,prl,12pt]{revtex4-2}   
\usepackage{graphicx}
\usepackage{amsmath}     
\usepackage{amssymb}
\usepackage{xcolor}
\usepackage{bm}
\usepackage{hyperref}

\hypersetup{colorlinks=true, linkcolor = [rgb]{0,0.08,0.45}, citecolor = [rgb]{0,0.08,0.45}, urlcolor = [rgb]{0,0.08,0.45}}

\newcommand{\half}{{\textstyle{\frac{1}{2}}}}

\begin{document}

\title{Regge spectral generator and form factors from hard exclusive amplitudes in holographic QCD}

\author{Guy~F.~de~T\'eramond}
\email[]{guy.deteramond@ucr.ac.cr}
\affiliation{Laboratorio de F\'isica Te\'orica y Computacional, 
Universidad de Costa Rica, 11501 San Jos\'e, Costa Rica}

\author{Stanley~J.~Brodsky}
\email[]{sjbth@slac.stanford.edu}
\affiliation{SLAC National Accelerator Laboratory, Stanford University, 
Stanford, California 94309, USA}

\author{Hans~G\"unter~Dosch}
\email{h.g.dosch@gmail.com}
\affiliation{Institut f\"ur Theoretische Physik der Universit\"at, 
D-69120 Heidelberg, Germany}

\date{\today}

\begin{abstract}

\vspace{40pt}

We show that the infinite tower of hard exclusive amplitudes in holographic
light-front QCD leads to a spectral generator $G(\alpha,\lambda)$ which encodes
the full Regge spectrum. The construction assumes a Poisson distribution of
Fock-state components, where $\lambda$ represents the average parton
multiplicity above the valence configuration. The resulting generator yields a
Regge spectrum invariant under continuous $\lambda$-deformations.
Its even- and odd-twist projections yield the meson and baryon form
factors with a single value of $\lambda$, fixed by the pion charge radius,
which in turn determines the full $Q^2$ dependence. The constituent counting
rules for both sectors follow from the structure of the spectral generator,
which also provides a compact analytic representation of parton distributions.

\end{abstract}

\maketitle

\newpage

\paragraph{Introduction.}

Unlike deep inelastic scattering, where the proton is typically
dissociated through incoherent processes, exclusive reactions at large
virtuality leave the target hadron intact~\cite{Lepage:1980fj}. At high $Q^2$
the virtual probe resolves short-distance structure inside the hadron,
providing a high-resolution view of its internal dynamics, while the scattering
amplitude remains coherent and preserves the bound state.
Exclusive amplitudes at large $Q^2$ obey the QCD constituent counting
rules~\cite{Brodsky:1973kr, Matveev:1972gb}. The same power-law behavior also
emerges in the gauge/gravity framework, where Polchinski and Strassler showed
that hard scattering in a confining background reproduces the twist scaling
expected from QCD~\cite{Polchinski:2001tt}.

This observation laid the foundation for holographic light-front
QCD~\cite{Brodsky:2006uqa,deTeramond:2008ht},  which rests on two
complementary routes: the mapping of the Drell--Yan--West form 
factor~\cite{Drell:1969km,West:1970av}---the
light-front overlap of hadronic wave functions summed over Fock
components---onto the equivalent bulk-to-boundary expression in
anti--de Sitter (AdS) space~\cite{Brodsky:2006uqa}, and the mapping of the semiclassical
light-front Hamiltonian eigenvalue equation onto the AdS wave equation, from
which the Regge spectrum emerges~\cite{deTeramond:2008ht}.
Exclusive processes are described in this framework in terms of coherent
contributions from the Fock-state components of the hadron, allowing
holographic methods to be applied even at large virtuality, where each twist
amplitude exhibits hard scaling at high $Q^2$~\cite{deTeramond:2025mls};
however, the coefficients of the Fock expansion have so far required
phenomenological input~\cite{Brodsky:2014yha}.

Motivated by the coherent structure of the light-front Fock expansion for exclusive
processes at arbitrary virtuality, we introduce in this Letter a
Poisson sum over the infinite tower of twist amplitudes. It defines
a closed-form Regge spectral generator $G(\alpha,\lambda)$, where $\alpha$ 
is the Regge trajectory  at fixed spin $S$ of the external current
and $\lambda$ represents the average parton multiplicity above the valence
configuration. The function $G(\alpha,\lambda)$ is analytic in $\alpha$, except for simple poles, 
which generate the full Regge spectrum for arbitrary spin. Its
Mittag--Leffler expansion shows that the pole positions in the spectral sums
are independent of $\lambda$, which enters only through the residues, implying that the Regge
spectrum is invariant under continuous $\lambda$-deformations.

The  even- and odd-twist projections of the spectral generator
yield the form factors and parton distributions
of mesons and baryons, respectively,
reflecting their $q\bar q$ and $qqq$ valence content. Both sectors share the
same Regge poles of the exchanged trajectory and reproduce the leading
constituent counting rules at large $Q^2$ through explicit residue sum rules, 
with a single Poisson parameter
$\lambda$ characterizing the probe. The meson/baryon distinction is thus
determined by the coupling to even or odd twist states of a common exchange
tower, with the pion form factor providing a quantitative test.

\vspace{10pt}

\paragraph{Holographic light-front QCD and Regge trajectories.}

In holographic light-front QCD (HLFQCD), the hadron spectrum is obtained
from a semiclassical approximation to the light-front invariant 
Hamiltonian equation $P^2 = M^2 \ge 0$.  This leads to frame-independent
relativistic wave equations and
form factors with the same structure as their counterparts in AdS space, thereby
establishing a  one-to-one correspondence between the light-front
Hamiltonian formalism and a gravity dual 
description~\cite{Brodsky:2006uqa,deTeramond:2008ht}.

Superconformal symmetry, a graded extension of conformal
symmetry~\cite{deAlfaro:1976vlx,Fubini:1984hf}, provides a natural
mechanism for the emergence of a confinement scale in HLFQCD. This is
particularly significant since QCD, in the limit of massless quarks, is
classically conformal and contains no intrinsic scale. The superconformal
construction determines the form of the effective interaction potential,
including nontrivial constant terms, and leads to hadronic spectra 
with $M^2 \sim n + L$, where $n$ and $L$ denote the radial and orbital 
quantum numbers. It also predicts supersymmetric relations linking 
mesons and baryons, with the pion emerging as a massless state in the 
chiral limit, thereby breaking hadronic supersymmetry~\cite{Dosch:2015nwa,deTeramond:2014asa}.

In holographic light-front QCD, Regge trajectories emerge dynamically
as solutions of the holographic light-front wave equations,
rather than being introduced phenomenologically.
Mesons and baryons organize along
linear Regge trajectories, $\alpha(s)=\alpha(0)+\alpha' s$, with their mass
spectrum determined by the on-shell condition
\[
\alpha(s=M^2)=J.
\]

For mesons, the total angular momentum is $J=L+S$, where $S$ is the quark spin,
and the Regge trajectory is given by
\begin{equation}
\alpha(s)= \left(\frac{S}{2}  - n - \frac{\Delta M^2}{4 \kappa^2} \right)+\alpha' s,
\qquad \alpha'=\frac{1}{4\kappa^2},
\label{alM}
\end{equation}
for fixed quantum number $n$.

The Regge slope $\alpha'$ is universal, a direct consequence of the
underlying superconformal structure, whereas the intercept $\alpha(0)$
differs for mesons and baryons and is lowered by the quark-mass
correction $\Delta M^2$~\cite{Dosch:2025lwb}.
For the light-quark vector trajectory this shift is small but not
negligible at the present level of precision: the analysis of
Ref.~\cite{Sufian:2018cpj} gives $\alpha(0) = 0.483$ for the $\rho$
trajectory, together with the spectral value $\kappa = 0.534$ GeV, the
inputs used in the determination of the Poisson parameter $\lambda$ below.

\vspace{10pt}

\paragraph{Form factors in holographic light-front QCD.}

In the gauge/gravity duality framework, hadronic form factors are obtained from 
the overlap of a conserved current propagating in AdS
space and normalizable bound-state wave 
functions~\cite{Polchinski:2002jw}. Holographic light-front QCD 
relates the twist dimension $\tau$, corresponding to the number of constituents 
in a given Fock-state component of the hadron, to the power-law behavior of the AdS
bound-state wave function, $\psi_\tau(z) \sim z^\tau e^{- \kappa^2 z^2/2}$, near the 
Minkowski conformal boundary. The resulting 
electromagnetic form factor for a spin-one external current and an arbitrary 
twist-$\tau$ Fock component of the target hadron can be written as~\cite{Brodsky:2014yha}
\begin{equation} \label{FFz}
F_\tau(t) \sim \int \frac{dz}{z^3} \psi_\tau(z) V(t, z) \psi_\tau(z) ,
\end{equation}
where $z$ is the fifth-dimensional variable of AdS space, the holographic variable, 
$V(t, z)$ is the AdS current and $t = - Q^2$, the momentum transfer variable.
Performing the $z$-integration in~(\ref{FFz}) using the integral 
representation of $V(t,z)$~\cite{Grigoryan:2007my} 
leads to~\cite{Zou:2018eam, deTeramond:2018ecg} 
\begin{equation} \label{FB}
F_\tau(t) \sim \int_0^1 (1-x)^{\tau-2} \,x ^{-\alpha' t} dx
= B(\tau -1, 1 -\alpha' t) ,
\end{equation}
where  $\alpha' = 1/ 4 \kappa^2$.

Our final expression for the twist-$\tau$ component of the form factor
\begin{equation} 
F_\tau(t) \propto B(\tau-1, S-\alpha(t)), 
\label{FFB}
\end{equation}
incorporates the spin $S$ of the external current and  $\alpha(t)$, the relevant Regge trajectory in the $t$-channel 
exchange. For the electromagnetic  form factor of the proton, 
for example, $S=1$, $\alpha(t)$ corresponds to a vector meson trajectory, 
and $\tau$ is the number of quarks probed by the external current in a given Fock-state 
component of the struck proton~\cite{deTeramond:2018ecg}.  For the gravitational 
form factor $A(t)$, $S=2$ and the Regge trajectory $\alpha_P(t)$ corresponds 
to the Pomeron~\cite{deTeramond:2021lxc}.  Equation~\eqref{FFB} is written
in terms of the full Regge trajectory, $\alpha(t)=\alpha(0)+\alpha' t$, thereby
incorporating the Regge intercept $\alpha(0)$, which is absent in
soft-wall models~\cite{Karch:2006pv}; its inclusion shifts the poles in the
form factor to their physical locations~\cite{Brodsky:2014yha}.

A similar expression was introduced soon after the development of the
Veneziano model, itself constructed to realize the duality
between resonances and Regge behavior~\cite{Dolen:1967jr}, by replacing
the $s$-channel dependence with a fixed
pole~\cite{Veneziano:1968yb,Ademollo:1969wd,Landshoff:1970ce,Bender:1969en,Bender:1970ew}:
\begin{equation}
F_\gamma(t) \propto B(\gamma, 1-\alpha(t)) \;\to\; |t|^{-\gamma},
\end{equation}
where $\gamma$ is determined by the decrease rate of the form factor
at large $|t|$. However, prior to the establishment of the constituent
counting rules in QCD~\cite{Brodsky:1973kr,Matveev:1972gb}, no relation between
integer values of $\gamma$ and the number of quark constituents could be established.

In holographic light-front QCD $\gamma$ is identified with the twist
through $\gamma = \tau - 1$~\cite{deTeramond:2018ecg} in the Fock
expansion. For integer twist $\tau=N\ge 2$, Eq.~\eqref{FFB} can
be written as a multipole amplitude, the product of $\tau-1$ simple
poles in the Regge variable,
\begin{equation} \label{FFprod}
F_\tau(t) \sim \prod_{n=0}^{\tau-2} \frac{1}{n+S-\alpha(t)} .
\end{equation}
This expression exhibits the characteristic twist scaling $F_\tau(t)\sim 1/t^{\tau-1}$ 
at large $t$, with pole positions at
\[
\alpha = S, S+1, \dots,  S+\tau-2 .
\]
Equivalently, Eq.~\eqref{FFprod} can be written in terms of the radial mass spectrum
of the probe by using the condition $\alpha(t=M_n^2)=S+n$,  for fixed $S$. We find
\begin{equation} \label{FFpoles}
F_\tau(t)=   \prod_{n=0}^{\tau-2}\frac{M_n^2}{M_n^2-t},
\end{equation}
with $F_\tau(0)=1$, and the spectral relation
\begin{equation} \label{Reggespec}
M_n^2=\frac{1}{\alpha'}\big(n+S-\alpha(0)\big),
\end{equation}
where $\alpha(0) = \frac{S}{2}  -  \frac{\Delta M^2}{4 \kappa^2}$ is the leading $(n=0)$ intercept.
The full hadron form factor is the coherent sum over the infinite
number of Fock components,
\begin{equation}
F_H(t) = \sum_{\tau} C_\tau \prod_{n=0}^{\tau-2}\frac{M_n^2}{M_n^2-t},
\label{FFH}
\end{equation}
where the normalization $F_H(0)=1$ requires $\sum_\tau C_\tau = 1$ if all
possible states are included.

\vspace{10pt}

\paragraph{Regge spectral generator.}

The constituent counting rules predict the power-law fall-off of amplitudes for
exclusive scattering processes according to the number of constituents that must
share the large momentum transfer in order to keep the hadron
intact~\cite{Brodsky:1973kr,Lepage:1980fj,Matveev:1972gb}.
The full exclusive amplitude therefore receives contributions from the entire
tower of twist components in the Fock basis, each contributing a
product of $\tau-1$ poles with hard scaling $f_\tau \sim 1/t^{\tau-1}$ for
large $t$.

We define a spectral generator $G(\alpha,\lambda)$ as the Poisson-weighted sum
of the infinite tower of twist amplitudes $f_\tau(\alpha)$,
\begin{equation} \label{Gsum}
G(\alpha,\lambda)  = \sum_{\tau=2}^{\infty}
\frac{\lambda^{\tau-2} e^{-\lambda}}{(\tau-2)!}\, B(\tau-1,S-\alpha),
\end{equation}
with $f_\tau(\alpha)\equiv B(\tau-1,S-\alpha)$.
The parameter $\lambda$ is the mean of the Poisson distribution
$P_n(\lambda)={\lambda^n e^{-\lambda}}/{n!}$, the average number of partons
above the minimal valence configuration, thus
$\langle \tau \rangle = 2 + \lambda$.
The Poisson generator $G(\alpha,\lambda)$ can be
expressed in closed form in terms of the confluent hypergeometric function
${}_1F_1$
\begin{align} \label{Grslt}
G(\alpha,\lambda)
& = e^{-\lambda} \! \int_0^1 (1-x)^{S - \alpha -1}  e^{\lambda x} dx  \nonumber \\
&= \frac{e^{-\lambda}}{S-\alpha} \, {}_1F_1(1; S-\alpha+1; \lambda),
\end{align}
as shown in the Appendix.

In the Drell--Yan--West frame the current does not couple to states with different 
particle number~\cite{Drell:1969km,West:1970av,Brodsky:1997de}. As a result 
the elastic form factor is a sum over Fock
components of definite constituent number, each contributing
independently. Likewise, in
holographic light-front QCD~\cite{deTeramond:2008ht,Brodsky:2014yha} the
semiclassical wave equations describe independent oscillator modes: their
coherent superposition should carry Poisson-distributed occupation
numbers, in direct analogy with the driven harmonic oscillator.

\vspace{10pt}

\paragraph{Mittag--Leffler expansion.}

The Regge spectrum generated by $G(\alpha,\lambda)$ becomes explicit through its
Mittag--Leffler expansion in the complex $\alpha$ plane. In fact,
Eq.~\eqref{Grslt} shows that the spectral generator is a meromorphic function
of $\alpha$ with simple poles. It admits the Mittag--Leffler decomposition
(see Appendix)
\begin{equation}
G(\alpha,\lambda)=
\sum_{n=0}^{\infty}
\frac{R_n(\lambda)}{n + S -\alpha},   \quad  R_n(\lambda)= \frac{(-\lambda)^n}{n!},
\label{MLexpal}
\end{equation}
where the residues $R_n(\lambda)$ are determined by the Poisson parameter $\lambda$.
Equation~\eqref{MLexpal} exhibits the Veneziano-like structure of an infinite
sequence of equidistant zero-width simple poles,
$\alpha=S+n$ with  $n=0,1,2,\dots $,
which are independent of $\lambda$: the confinement spectrum is invariant
under continuous $\lambda$-deformations of the Poisson distribution.

Structurally, the Poisson generator~\eqref{Gsum} reorganizes the tower of
fixed-twist amplitudes, each a finite product of poles, into an infinite sum
of simple Regge poles~\eqref{MLexpal}, thereby providing an analytic mechanism
through which the confining structure of individual Fock components generates
the full Regge mass spectrum of radial excitations~\eqref{Reggespec} via the
spectral condition $\alpha(M_n^2)=S+n$.

\vspace{10pt}

\paragraph{Even-odd projection of the spectral generator. \label{EOG} } 
As the virtual photon propagates through the confinement domain at distances
$\sim 1/\kappa$, it undergoes strong interactions and fluctuates into
intermediate vector states: the current hadronizes into $q\bar q$ pairs and
their radial excitations. This dressing of the current is encoded in the
universal spectral generator $G(\alpha,\lambda)$, which contains all the
states of the Regge trajectory in the exchange channel, Eq.~\eqref{MLexpal},
independently of the target hadron. The target selects the components
matching its own light-front Fock expansion, leading to a distinct effective
coupling for each hadron class, even though the coupling of the
electromagnetic current at the quark level is universal.

In the semiclassical approximation, gluonic degrees of freedom are sublimated
into the effective confining potential and do not appear as
constituents~\cite{Brodsky:2011pw}. The Poisson distribution weights
 each twist component of the dressed current;
the target singles out even twists for a meson, whose Fock tower
$|q\bar q\rangle, |q\bar q\, q\bar q\rangle, \ldots$ carries
$\tau = 2, 4, \ldots$, and odd twists for a baryon, with
$|qqq\rangle, |qqq\, q\bar q\rangle, \ldots$ and $\tau = 3, 5, \ldots$.

We consider first the case where the probe couples to the even twist states
of a target meson (see Appendix)
\begin{align}
G_E(\alpha,\lambda)&=\sum_{\substack{\tau=2 \\ \tau\,\text{even}}}^{\infty}
\frac{\lambda^{\tau-2} e^{-\lambda}}{(\tau-2)!}\,
B(\tau-1,S-\alpha)  \nonumber \\
&= e^{-\lambda}
\int_0^1 (1-x)^{S-\alpha-1} \cosh(\lambda x)\,dx \nonumber \\
&= \frac{e^{-\lambda}}{2(S-\alpha)}
\left[{}_1F_1\!\big(1;\,S-\alpha+1;\,\lambda\big)
+ {}_1F_1\!\big(1;\,S-\alpha+1;\,-\lambda\big)\right].
\label{GE}
\end{align}
Likewise, for the coupling of the probe to the odd twist states of a target
baryon we find
\begin{align}
G_O(\alpha,\lambda)&=\sum_{\substack{\tau=3 \\ \tau\,\text{odd}}}^{\infty}
\frac{\lambda^{\tau-2} e^{-\lambda}}{(\tau-2)!}\,
B(\tau-1,S-\alpha)  \nonumber \\
&= e^{-\lambda}
\int_0^1 (1-x)^{S-\alpha-1} \sinh(\lambda x)\,dx \nonumber \\
&= \frac{e^{-\lambda}}{2(S-\alpha)}
\left[{}_1F_1\!\big(1;\,S-\alpha+1;\,\lambda\big)
- {}_1F_1\!\big(1;\,S-\alpha+1;\,-\lambda\big)\right],
\label{GO}
\end{align}
with $G(\alpha, \lambda) = G_E(\alpha, \lambda) + G_O(\alpha, \lambda)$.

Since ${}_1F_1$ is an entire function of its argument $\lambda$,
and the sign of $\lambda$ enters~\eqref{GE} and \eqref{GO} only through this
argument, without modifying the $S-\alpha$ dependence, both projections
inherit the full pole ladder of the Mittag--Leffler
expansion~\eqref{MLexpal}. Mesons and baryons couple to the
complete tower of states of the exchanged current, $\alpha=S+n$, with
distinct couplings given by the residues $R_n^{E,O}$ (see Appendix)
\begin{equation} \label{RnEO}
G_{E,O}(\alpha,\lambda)=\sum_{n=0}^{\infty}\frac{R_n^{E,O}(\lambda)}{n + S - \alpha}, \quad
R_n^{E,O}(\lambda)=\frac{\lambda^n}{2\,n!}
\left[(-1)^n \pm e^{-2\lambda}\right],
\end{equation}
with $R_n(\lambda) = R_n^E(\lambda) + R_n^O(\lambda)$.
The residues obey the sum rules
\begin{equation} \label{sumrules}
\sum_{n=0}^{\infty} R_n^{E}(\lambda)=e^{-\lambda},
\qquad
\sum_{n=0}^{\infty} R_n^{O}(\lambda)=0,
\end{equation}
which determine the large-$|S-\alpha|$ behavior of the two sectors:
\begin{equation} \label{GEOasymp}
G_E(\alpha,\lambda) \sim \frac{e^{-\lambda}}{S-\alpha},
\qquad
G_O(\alpha,\lambda) \sim \frac{\lambda\, e^{-\lambda}}{(S-\alpha)^{2}},
\end{equation}
since the vanishing residue sum of the odd projection cancels its leading
power.

\vspace{10pt}

\paragraph{Meson-baryon form factors.}

The even and odd projections of the Regge spectral generator determine the
meson and baryon form factors, $F_M(t,\lambda)$ and $F_B(t,\lambda)$, up to
an overall normalization:
\begin{equation}
F_{M,B}(t,\lambda) =
\frac{G_{E,O}(\alpha(t),\lambda)}{G_{E,O}(\alpha(0),\lambda)},
\label{GFt}
\end{equation}
where $F_{M,B}(0,\lambda)=1$. In the Regge limit, $\alpha(t)\sim\alpha' t$
at large space-like $t$, the asymptotic forms~\eqref{GEOasymp} reproduce the
leading constituent counting rules~\cite{Brodsky:1973kr,Matveev:1972gb}
\begin{equation}
F_M \sim \frac{1}{\alpha' |t|},
\qquad
F_B \sim \frac{1}{(\alpha' |t|)^{2}},
\label{FMFBasy}
\end{equation}
through the explicit residue sum rules~\eqref{sumrules}. This behavior is
also consistent with the Phragm\'en--Lindel\"of theorem, which relates the
asymptotic limits in the space-like and time-like regions,
$F(t\to-\infty)\sim F(s\to+\infty)$, up to a phase~\cite{Pacetti:2014jai}.

\begin{figure}[h!]
    \centering 
    \includegraphics[width=0.56\linewidth]{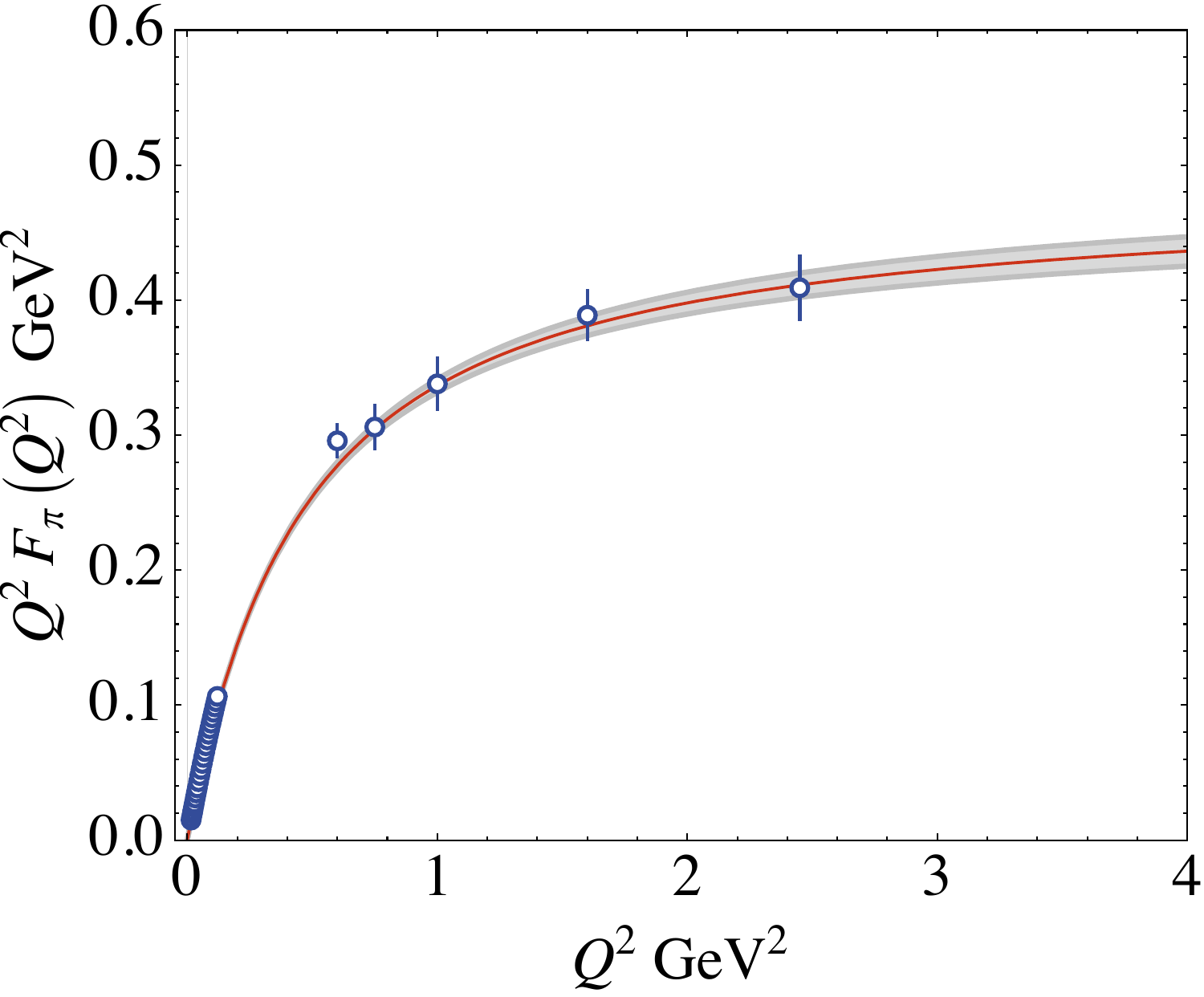}
    \caption{Experimental data for the pion electromagnetic form factor in
the space-like ($Q^2 \ge 0$) region are compared with the model prediction
(red curve) for $\lambda = 0.87 \pm 0.07$, with the uncertainty band in
grey. Data from NA7~\cite{NA7:1986vav} and
JLab~\cite{Horn:2007ug,JeffersonLab:2008jve}.
\label{fig:pionEMFF}} 
\end{figure}

We determine the mean parton excitation above the valence configuration,
$\lambda$, from the measured pion charge radius $\langle r_\pi^2\rangle$
and $F_M(t,\lambda)$, using the vector meson Regge trajectory~\eqref{alM}
for $S=1$ and the spectral values $\kappa = 0.534$ GeV and $\alpha(0)_\rho = 0.483$
from~\cite{Sufian:2018cpj}. We obtain $\lambda = 0.87 \pm 0.07$ from the PDG
value $\langle r_\pi^2\rangle = 0.434 \pm 0.005~\mathrm{fm}^2$~\cite{ParticleDataGroup:2024cfk}. 
Having fixed $\lambda$,  we compute the space-like $Q^2 = - t \ge 0$ dependence of the pion form
factor as a parameter-free prediction, shown in Fig.~\ref{fig:pionEMFF}.

The compact expression of the form factors~\eqref{GFt} can be represented
as a sum of products of poles by using the spectral
relation~\eqref{Reggespec}. We find
\begin{align}
F_M(t,\lambda) &= \frac{1}{N_M}
\sum_{n=0}^{\infty}
\frac{\lambda^{2n}}{(\alpha')^{2n+1}}
\prod_{k=0}^{2n} \frac{1}{M_k^2 - t}, \label{FME}\\
F_B(t,\lambda) &= \frac{1}{N_B}
\sum_{n=0}^{\infty}
\frac{\lambda^{2n+1}}{(\alpha')^{2n+2}}
\prod_{k=0}^{2n+1} \frac{1}{M_k^2 - t}, \label{FBE}
\end{align}
where $N_M$ and $N_B$ are normalization factors. In practice, truncating
the sum to a few terms already yields a result nearly indistinguishable
from the compact expression~\eqref{GFt}. 
The expanded form is useful to extend the model to the time-like domain,
$t>0$, where resonance widths and the phase space of the opening decay
channels require phenomenological input or a specific model computation.

The expanded form is also useful for comparison with previous
holographic results. From the meson form factor~\eqref{FME} we obtain the
twist coefficients $C_\tau$ in~\eqref{FFH},
\begin{equation}
C_2 = 0.83, \qquad C_4 = 0.16, \qquad C_6 = 0.008, \qquad
C_8 \simeq 2\times 10^{-4},
\label{Ctau}
\end{equation}
showing the rapid convergence of the twist expansion. The coefficients,
previously treated as fitted parameters~\cite{Brodsky:2014yha}, are here
determined to all orders, and the fast convergence of the truncated
expansion in~\cite{Brodsky:2014yha} follows as a structural consequence of
the spectral generator.

The coupling of the dressed probe to the orbital components of the
target, critical for the nucleon Pauli form factor and the neutron
sector~\cite{Brodsky:2014yha}, is not determined in the present Letter.
A detailed analysis of the nucleon form factors, which requires the
internal spin-flavor and orbital structure of the nucleon Fock components,
is left for future work.

\vspace{4pt}

\paragraph{Parton distributions.} 

Parton distributions in holographic light-front QCD are given
by~\cite{deTeramond:2018ecg,deTeramond:2021lxc}
\begin{equation}
h_\tau(\alpha,x)=w(x)^{S-\alpha-1}\,w'(x)\left(1-w(x)\right)^{\tau-2},
\end{equation}
where $\int_0^1 h_\tau(\alpha,x)\,dx = B(\tau-1,S-\alpha)$ independently of
the specific form of $w(x)$. The function $w(x)$ is assumed to be identical for
quarks and gluons and is constrained by imposing the QCD inclusive counting
rules at $x \to 1$, namely $h_\tau \sim(1-x)^{2\tau-3}$~\cite{Drell:1969km,Blankenbecler:1974tm},
as well as Regge behavior at small $x$, $h_\tau \sim x^{S - \alpha - 1}$. This leads to the conditions $w'(1)=0$
and $w(x)\sim x$ as $x\to 0$~\cite{deTeramond:2018ecg}. A natural question is how this structure is modified
by the coherent Poisson summation over the full tower of twist amplitudes.

To examine this point we define, in analogy with the spectral generator,
the even and odd sums of the twist tower of parton distributions, $K_E$
and $K_O$:
\begin{align}
K_E(\alpha,x,\lambda) &= \sum_{\substack{\tau=2 \\ \tau\,\text{even}}}^{\infty}
\frac{\lambda^{\tau-2} e^{-\lambda}}{(\tau-2)!}\, h_\tau(\alpha,x)
\nonumber \\
&= e^{-\lambda}\, w(x)^{S-\alpha-1}\, w'(x)
\cosh\big(\lambda\,(1-w(x))\big),
\label{KE}
\end{align}
\begin{align}
K_O(\alpha,x,\lambda) &= \sum_{\substack{\tau=3 \\ \tau\,\text{odd}}}^{\infty}
\frac{\lambda^{\tau-2} e^{-\lambda}}{(\tau-2)!}\, h_\tau(\alpha,x)
\nonumber \\
&= e^{-\lambda}\, w(x)^{S-\alpha-1}\, w'(x)\,
\sinh\big(\lambda\,(1-w(x))\big),
\label{KO}
\end{align}
with $\int_0^1 K_{E,O}(\alpha,x,\lambda)\,dx = G_{E,O}(\alpha,\lambda)$.

Near $x=1$, the conditions $w(1)=1$ and $w'(1)=0$ give the
model-independent expansion $1-w(x) = -\tfrac{1}{2}\,w''(1)(1-x)^2$,
with $w''(1)<0$, and $w'(x) = - w''(1)(1-x)$. Expanding the hyperbolic 
kernels at the endpoint, the projections~\eqref{KE} 
and \eqref{KO} reduce to their minimal-twist components,
\begin{equation}
K_E \sim e^{-\lambda}\,(1-x),
\qquad
K_O \sim \lambda\, e^{-\lambda}\,(1-x)^{3},
\label{KEOlimits}
\end{equation}
reproducing the inclusive counting rules $(1-x)^{2\tau_{\min}-3}$ for
$\tau_{\min}=2$ and $3$~\cite{Drell:1969km,Blankenbecler:1974tm}, with the
Poisson weights $e^{-\lambda}$ and $\lambda e^{-\lambda}$ of the valence
$q\bar q$ and $qqq$ configurations. At small $x$, where $w(x)\sim x$, both
projections retain the Regge behavior $\sim x^{S-\alpha-1}$.

The physical parton distributions are obtained by normalizing the projected
sums at $t=0$,
\begin{equation}
\mathcal{H}_{M,B}(x,t,\lambda) =
\frac{K_{E,O}(\alpha(t),x,\lambda)}{G_{E,O}(\alpha(0),\lambda)},
\label{HMB}
\end{equation}
with $\int_0^1 \mathcal{H}_{M,B}(x,0,\lambda)\,dx = 1$. At finite momentum
transfer the $x$-integral reproduces the form factors~\eqref{GFt},
$\int_0^1 \mathcal{H}_{M,B}(x,t,\lambda)\,dx = F_{M,B}(t,\lambda)$, since
the projected distributions and form factors derive from the same spectral
structure.

\vspace{8pt}

\paragraph{Concluding remarks.}

The structural analysis presented here organizes the full twist tower of
exclusive amplitudes into a single generating function, whose physical
consequences---the Regge spectrum, the meson/baryon projection, the
constituent counting rules, the form factors and parton
distributions---follow from the universal spectral characteristics of the
probe and the meson/baryon nature of the target. The pion form factor
provides a quantitative test of this framework.

The spectral generator $G(\alpha,\lambda)$ is the Poisson-weighted coherent
sum of the infinite tower of twist amplitudes, evaluated in closed form as
a meromorphic function of the Regge trajectory $\alpha$. Its
Mittag--Leffler expansion exhibits simple poles at $\alpha = S+n$,
corresponding to the radial excitations of the exchanged trajectory, with
positions independent of $\lambda$.

The even- and odd-twist projections
of the Poisson generator yield the meson and baryon form factors and parton
distributions, reflecting the $q\bar q$ and $qqq$ valence content of the
two hadron sectors. Both projections share the Regge poles of the common
exchanged trajectory, and the constituent counting
rules follow from the residue sum
rules of the spectral generator: $\sum_n R_n^{E} = e^{-\lambda}$ fixes the
leading meson power, while the vanishing sum $\sum_n R_n^{O} = 0$ produces
the additional baryon suppression. The exclusive counting rules in $t$ and
the inclusive counting rules at $x\to 1$ are both governed by identical Poisson
weights of the minimal Fock configuration of each sector, with the single parameter 
$\lambda$ fixed by the pion charge radius. The extension to the nucleon sector, which
involves its internal spin-flavor and orbital structure, is deferred to future work. The analytic
framework introduced in the present work can be extended to other 
exclusive processes including form factors and parton distributions 
of hadrons with heavy and light quarks.

The present framework connects two historically distinct strands: the
duality program of the pre-QCD era, in which global spectral information
was used to constrain local
amplitudes~\cite{Dolen:1967jr,Veneziano:1968yb,Bender:1970ew}, and the
constituent counting rules of perturbative QCD ~\cite{Brodsky:1973kr,Matveev:1972gb},
which determine the local
short-distance behavior. In the spectral generator both aspects reside in
a single analytic function: its global residue structure, summed over the
full nonperturbative tower, determines the local power-law behavior
through exact sum rules. The coherent hard exclusive amplitude
is thereby determined within this framework by
the confinement spectrum itself.

Exact analytic Regge spectral sums have recently been obtained in
QCD(1+1)~\cite{Artemev:2025cev} using nonperturbative methods developed
in~\cite{Fateev:2009jf}. A corresponding derivation from first-principles
QCD in (3+1) dimensions remains an outstanding challenge.

\vspace{8pt}

\paragraph{Acknowledgments.} 

The authors are grateful to Tanja Horn for her guidance regarding the JLab pion form factor experimental data.


\appendix

\section{Appendix: Some useful results \label{app}}

\paragraph{Integral representations.}

Two integral representations are used in the derivations below.
The Euler Beta function $B(u,v) = \Gamma(u)\Gamma(v)/\Gamma(u+v)$ admits
the representation
\begin{equation}
B(u,v)=\int_0^1 t^{u-1}(1-t)^{v-1}\,dt,
\label{Betaint}
\end{equation}
with $\Re(u)>0$, $\Re(v)>0$. The confluent hypergeometric function
${}_1F_1(a;b;z)$, defined by the series
\begin{equation}
{}_1F_1(a;b;z)=
\frac{\Gamma(b)}{\Gamma(a)}
\sum_{n=0}^{\infty}
\frac{\Gamma(a+n)}{\Gamma(b+n)}
\frac{z^n}{n!},
\label{1F1ser}
\end{equation}
admits the integral representation
\begin{equation}
{}_1F_1(a;b;z)=
\frac{\Gamma(b)}{\Gamma(a)\,\Gamma(b-a)}
\int_0^1 e^{zt}\,t^{a-1}(1-t)^{b-a-1}\,dt,
\label{1F1int}
\end{equation}
valid for $\Re(b)>\Re(a)>0$.
 
\vspace{10pt}  

\paragraph{Poisson generating function.}

Consider the Poisson generating function $G(u,s)$ for the
Beta function sequence $f_n(u) = B(n+1,u)$,
\begin{equation} \label{Gus}
G(u,s)=\sum_{n=0}^{\infty} \frac{s^n e^{-s}}{n!}\,B(n+1,u),
\end{equation}
where $s$ is the mean of the Poisson probability distribution
$P_n(s)={s^n e^{-s}}/{n!}$, with $\langle n\rangle = s$.

Using the integral representation~\eqref{Betaint} and interchanging
the sum and integral, we obtain
\begin{equation} \label{GusInt}
G(u,s)= e^{-s}\int_0^1 (1-t)^{u-1}\,e^{st}\,dt.
\end{equation}
Applying~\eqref{1F1int} with $a=1$, $b=u+1$, and $z=s$ then gives
\begin{equation} \label{G1F1}
G(u,s)=\frac{e^{-s}}{u}\,{}_1F_1(1;\,u+1;\,s).
\end{equation}

\vspace{10pt}

\paragraph{Mittag--Leffler expansion.}

The pole expansion of $G(u,s)$ follows from Kummer's transformation~\cite{abramowitz1965handbook}
\begin{equation}
e^{-z}\,{}_1F_1(a;\,b;\,z) = {}_1F_1(b-a;\,b;\,-z).
\label{Kummer}
\end{equation}
Applying~\eqref{Kummer} with $a=1$, $b=u+1$, $z=\pm s$, and expanding the
right-hand side with the series~\eqref{1F1ser}, we find
\begin{equation}
\frac{1}{u}\,{}_1F_1(1;u+1;\pm s) 
= e^{\pm s} \sum_{n=0}^{\infty} \frac{(\mp s)^n}{n!\,(u+n)}.
\label{G1F1K}
\end{equation} 
The Poisson generator~\eqref{G1F1} then admits the Mittag--Leffler expansion
\begin{equation} \label{MLexp}
G(u,s)=\sum_{n=0}^{\infty}\frac{R_n(s)}{u+n}, \qquad R_n(s)=\frac{(-s)^n}{n!},
\end{equation}
with simple poles at $u = -n$ ($n = 0, 1, 2, \ldots$).
 
\vspace{10pt}
 
\paragraph{Even-odd projection.}

The Poisson generator $G(u,s)$ admits a decomposition into even and odd
infinite sums while preserving its analytic properties. The even sum
\begin{equation} \label{GusE}
G_E(u,s) = \sum_{n=0}^\infty
\frac{s^{2n}\,e^{-s}}{(2n)!}\,B(2n+1,u)
\end{equation}
is evaluated using~\eqref{Betaint} and the series expansion of the
hyperbolic cosine, giving
\begin{equation} \label{GusEInt}
G_E(u,s)= e^{-s}\int_0^1 (1-t)^{u-1}\cosh(st)\,dt.
\end{equation}
Using $\cosh(st) = \half\!\left(e^{st}+e^{-st}\right)$ and~\eqref{GusInt},
we obtain
\begin{equation} \label{GEa}
G_E(u,s)= \frac{e^{-s}}{2u}
\Bigl[\,{}_1F_1(1;\,u+1;\,s)+{}_1F_1(1;\,u+1;\,-s)\Bigr].
\end{equation}
The odd sum is obtained by the same steps with the hyperbolic sine:
\begin{align}
G_O(u,s)
&= \sum_{n=0}^\infty\frac{s^{2n+1}\,e^{-s}}{(2n+1)!}\,B(2n+2,u)
\nonumber \\
&= e^{-s}\int_0^1 (1-t)^{u-1}\sinh(st)\,dt \nonumber \\
&= \frac{e^{-s}}{2u}
\Bigl[\,{}_1F_1(1;\,u+1;\,s)-{}_1F_1(1;\,u+1;\,-s)\Bigr],
\label{GOa}
\end{align}
with $G(u,s) = G_E(u,s)+G_O(u,s)$ exactly.

\vspace{10pt}

\paragraph{Residues of the even-odd projections.}

The Mittag--Leffler expansions of the even-odd projections follow from~\eqref{G1F1K}
\begin{equation} \label{MLEO}
G_{E,O}(u,s)=\sum_{n=0}^{\infty}\frac{R_n^{E,O}(s)}{u+n},
\qquad
R_n^{E,O}(s)=\frac{s^n}{2\,n!}\left[(-1)^n \pm e^{-2s}\right],
\end{equation}
with $R_n^{E}+R_n^{O}=(-s)^n/n!$, the residues of the full generator.
Both projections thus contain the complete pole ladder $u=-n$.
 The residues obey the sum rules
\begin{equation} \label{sumrulesA}
\sum_{n=0}^{\infty} R_n^{E}(s)=e^{-s},
\qquad
\sum_{n=0}^{\infty} R_n^{O}(s)=0 ,
\end{equation}
which control the large-$u$ behavior.  Expanding~\eqref{MLEO} for large $u$:
$u  \,G(u) \simeq \sum_n R_n - \frac{1}{u}\sum_n n\,R_n + \cdots,$  one  obtains
the asymptotic expressions
\begin{equation}
G_E(u,s) \sim \frac{e^{-s}}{u},  \qquad G_O(u,s) \sim \frac{e^{-s}\,s}{u^2}.
\label{GEOasympA}
\end{equation}
The vanishing residue sum of the odd projection cancels its leading power,
and the first moment, $\sum_n n\,R_n^{O} = -s\,e^{-s}$, fixes the coefficient
of the subleading term. For the even projection the first moment vanishes,
$\sum_n n\,R_n^{E}=0$, thus corrections to the leading behavior begin only at
order $1/u^{3}$.

The same asymptotic behavior follows independently from the
large-$u$ expansion 
\begin{equation}
{}_1F_1(1;\,u+1;\,\pm s)
= 1 \pm \frac{s}{u+1} + \mathcal{O}\!\left(\frac{1}{u^2}\right),
\label{1F1asymp}
\end{equation}
providing a
consistency check of the sum rules~\eqref{sumrulesA}.

Identifying $u=S-\alpha>0$ and $s=\lambda$ reproduces the
expressions used in the main text.

%

\end{thebibliography}


\end{document}